\documentclass[12pt]{article}
\usepackage{graphics}
\usepackage{epsfig}
\usepackage{hyperref}
\usepackage{graphics}
\usepackage{color}      
\usepackage{graphicx}

\title{\bf Noise-created bistability and stochastic resonance of impurities diffusing in 
a semiconductor layer }
\author{Mesfin Asfaw$^{a}$\thanks{ Electronic address:
mesfin.taye@csun.edu}, Berhanu Aragie$^{b}$ and Mulugeta Bekele$^{b}$ \\
     $^{a}$Department of Physics and Astronomy, California State University\\ Northridge, California,USA\\ $^b$ Department of Physics, Addis Ababa University\\
P.O.Box 1176, Addis Ababa, Ethiopia }
\date{Received: date / Revised version: date}

\begin{document}
\maketitle
\maketitle
\begin{abstract}
 
 We  investigate   the dynamics  of  impurities walking   along a semiconductor layer assisted by thermal noise of strength  $D$ and  external harmonic potential  $V(x)$.  Applying   a nonhomogeneous hot temperature   in the vicinity of the potential minimum   may modify the external potential into a bistable effective potential.
  We propose the ways of mobilizing and eradicating
the unwanted impurities along the semiconductor
 layer.  Furthermore, the thermally activated rate of hopping for  the impurities as a function of the model parameters  is studied in   high barrier limit.
Via two state approximation, we  also study the stochastic resonance  (SR) of the impurities  dynamics where the same noise source  that induces the
 dynamics also induces the transition from mono-stable to bistable state which leads to SR in the presence of time varying field.  
\end{abstract}

\section{Introduction}
Recently the physics of semiconductors has received considerable attentions  as devices
made from semiconductors are vital in the  construction of  modern electronic equipment. One of the
beneficial features of these  semiconductors is the possibility to adjust their conductivities by
introducing    impurities (dopants) into their crystal lattice.
The level of conductivity can be controlled by the amount  and type of impurities.
Unlike conductors, their conductivity increase with temperature; this property   makes them even
more important.

Since the conductivity of the semiconductor relies  not only on the concentration of the dopants but also on
the thermal background  strength of the medium,  the physics of    thermal diffusion   of the impurities along
the semiconductor layer has attracted considerable attentions \cite{s1,s7,s3,s4}. Recently, the  proposal by \cite{s1} unveiled   the method of eliminating
unwanted impurities from the region of the semiconductor via  movable  external harmonic potential  along the semiconductor. The same
group has  investigated  how to control the diffusion of impurities via an external potential which  has an advantage of preserving
the crystal structure as diffusion can  take place at low temperature \cite{s7}. Their numerical results exhibit that the diffusion  of the dopant increases when 
the strength of the external potential decreases. At high  doping density, the internal field  becomes significant and it renormalizes
 the effect of the  external potential. Thus the diffusivity of the dopant increases as the dopant density increases.

 Because manipulating  the diffusion of  impurities (either acceptors or donors) to a desired region of the semiconductor layer is
vital, in this work we propose different ways of mobilizing the impurities along the semiconductor. We  consider  the   impurities walking from  one lattice trap to the other lattice trap with  a trap depth potential  along the semiconductor layer assisted by thermal noise of strength  $D$ and the  external harmonic potential.      Furthermore,  a nonhomogeneous hot temperature  is applied  in the vicinity of the potential minimum which  may modify the external potential into a bistable effective potential.
 Neglecting the interaction between the dopants at  low  impurity density,  we show that the diffusion of impurities can be controlled not only by varying the external potential but also by altering the intensity of  the temperature  of the hot spot  and trap depth of the impurities. We also propose the way of eradicating the unwanted impurities from the certain region of the semiconductor by varying  the different model parameters  without considering movable external potential. Furthermore, we investigate the {dependence} of the rate on different model parameters.

One very crucial but unexplored issue is the way of enhancing the mobility of the impurities along  the semiconductor.
This can be achieved by applying
symmetry  breaking fields such  as  time varying   signal which may lead the system into stochastic resonance. Stochastic resonance has been widely studied over the last  few decades \cite{c12,  c15, c16, c17, re20} and  has been adopted
 to describe an interesting statistical property of periodically
  modulated and noise driven multi-stable dynamical system. It exhibits that
under proper condition an increase in the input noise level
results in an increase in the output signal-to-noise ratio.

The  phenomenon of SR  was first introduced by Benzi \cite{c11}. Later, the  idea of stochastic resonance has been implemented in many model systems. Notable examples include: SR for confined systems \cite{c21}, SR for complex systems such as  polymers  \cite{c15,c16,c22,cc22} and  SR for a  Brownian particle moving across a porous membrane \cite{cm22}. The  vast majority of studies on stochastic resonance have been focused on analyzing the dynamics of a bistable system. Among such
  models,  the two state model \cite{c12,c23}  has been proven to be extremely useful in the understanding of
 the stochastic phenomenon offering  a simple framework able to
 provide analytical results

 In this paper, via two state approximation, we  study the stochastic resonance  (SR) of a particular
 noise sustained dynamics where the same noise source  that induces the
 dynamics also induces the transition from mono-stable to bistable state which leads to SR in the presence of time varying field. 
We find that the spectral amplification  $\eta$ attains an optimal value at a certain finite value of  noise strength $D_{opt}$. As the temperature   of the hot spot  increases, the peak of $\eta$ decreases. On the other hand, the value of  thermal strength $D_{opt}$ increases as the intensity of the  nonhomogenous  hot temperature increases. The strength of trap depth also considerably affects the $\eta$. We find that the peak of the $\eta$ decreases as  the strength of the potential trap  increases. Furthermore, we explore the {dependence} of $\eta$ on other model parameters.

It is important to note that the results and model   presented in
this work are not specific to   the impurities dynamics. Rather
the present study serves as a basic paradigm in which to understand
diffusion and   noise induced nonequilibrum phase transition in 
discrete systems. Thus the present model is of broad interest in
various fields.

The rest of the paper is organized as follows: in Section 2, we
present the model. In Section 3, we study the diffusion of the impurities  in the presence of a 
nonhomogeneous temperature. In section 4, we study the SR of
impurities and explore how the  $\eta$ for the impurities behaves
as a function of the model parameters.
 Section 5 deals with summary and conclusion.

\section{The model }

We consider a one-dimensional system where non-interacting impurities jump from one
lattice trap to the other lattice trap  assisted by thermal noise and  external
 potential energy which is monostable
\begin{equation}
V(x)=V_{0}x^{2}
 \end{equation}
where $V_{0}$ and $x$ denote the  potential energy and the position
of the impurities, respectively. When  background temperature is homogeneous,
particles concentrate  around the potential minimum.  On the other
hand, in the presence of a nonhomogeneous temperature background,
\begin{equation}
T(x)=T_{c}+T_{h}exp[-\frac{x^{2}}{2\sigma^{2}}]
\end{equation}
which is hot around the potential minimum and decays  to a lower temperature  as one goes away on both sides from the potential minimum,
the system may undergo a phase transition in which the particles
pile up around two points  of the potential minima. Here
the  parameters $T_{h}$ and $T_{c}$  designate the temperature of
the  hot and   cold reservoirs, respectively, while $\sigma$
denotes the standard deviation.

At low impurity density,  the impurity dynamics is governed by \cite{c24,c25}
\begin{equation}
\frac{\partial P(x,t)}{\partial t}=\frac{\partial}{\partial
x}\{\frac{V^{'}(
x)}{k_{B}T(x)}exp[-\frac{\Phi}{k_{B}T(x)}]P(x,t)+\frac{\partial}{\partial
x}exp[-\frac{\Phi}{k_{B}T(x)}]P(x,t)\}
\end{equation}
where $P(x, t)$ is the impurity density at a position $x$ and time $t$. The impurities  jump   from one lattice trap, which has an internal potential depth (trap depth) $\Phi$, to the next lattice trap along the one
dimensional lattice.  
The steady state probability distribution of the impurity
$P_{ss}(x)$ is given by
\begin{equation}
P_{ss}(x)=Ce^{\frac{\Phi}{K_{B}T(x)}}exp[-\int^{x}_{0}\frac{V'(y)}{k_{B}T(y)}dy]
\end{equation}
where C is a normalizing constant. Rearranging Eq. 4,  one gets
the probability distribution
\begin{equation}
 P_{ss}(x)=C_{0}e^{-\frac{V_{eff}(x)}{k_{B}T_{c}}}
\end{equation}
with an effective potential  energy $V_{eff}(x)$ which is  given by
\begin{equation}
V_{eff}(x)=V_{0}x^{2}+2V_{0}\sigma^{2}ln[\frac{1+\alpha
e^{-\frac{x^{2}}{2\sigma^{2}}}}{1+\alpha}]-\frac{\Phi}{[1+\alpha
e^{-\frac{x^{2}}{2\sigma^{2}}}]}
 \end{equation}
 where dimensionless parameter $\alpha$  relates $T_{h}$ and $T_{c}$ as $
T_{h}=\alpha T_{c}$. The effective potential energy is bistable
when the condition
\begin{equation}
\Phi>2V_{0}\sigma^{2}\left({\alpha+1\over \alpha}\right)
\end{equation}
is met. In this regime, the location of the saddle point is  at
$x = 0$ while the two symmetric stable points are located at
\begin{equation}
x_{m}=\pm \sqrt{2}\sigma \sqrt{\ln[\alpha({\Phi\over 2\sigma^2V_{0}}-1)]}.
\end{equation}
The effective potential energies at the saddle and stable points are given as
\begin{equation}
V_{eff}(0)={-\Phi \over (1+\alpha)}
\end{equation}
and
\begin{equation}
V_{eff}(x_{m})=-\Phi +2\sigma^2V_{0}\left[1+\ln\left[{\Phi \alpha\over 2\sigma^{2}V_{0}}-\alpha\right]+\ln\left[{\Phi \over (1+\alpha)(\Phi-2\sigma^2V_{0})}\right]\right],
\end{equation}
respectively. The two minima of the potential are separated by a
barrier of height $\Delta V^{eff}=V_{eff}(0)-V_{eff}(x_{m})$. The
curvatures at the barrier top $\omega_{0}$ and the well minima
$\omega_{x_{m}}$ take  simple forms
\begin{equation}
\omega_{0}={\left({-\alpha \Phi /
\sigma^{2}}-2\alpha(1+\alpha)V_{0}+2(1+\alpha)^2V_{0}\right)\over (1+\alpha)^2},
\end{equation}
and
\begin{equation}
\omega_{x_{m}}={4V_{0}\over
\Phi^{2}}(\Phi-2\sigma^2V_{0})^2\ln[\alpha({\Phi\over
2\sigma^2V_{0}}-1)],
\end{equation}
respectively.

Transforming  Eq. (4) into  Eq. (5) is equivalent to converting the non-equilibrium problem to an equilibrium one
where the usual Boltzmann's statistics holds. Hereafter we consider impurities undergoing a random walk motion along the effective potential
assisted by the thermal kicks   $D=k_{B}T_{c}$ where $k_{B}$ is the Boltzmann constant. One can note that
the information about  the non-equilibrium features of the system is stored in the effective potential. 
 In the next section we explore how the dynamics of the impurities under the  effective potential $V^{eff}$   behaves as a function of the model parameters.

\section{Impurity diffusion in a nonhomogeneous temperature}

 We consider   non-interacting impurities (either acceptors or donors \cite {s7} ) of low density hopping from one lattice site to the other site within the semiconductor layer.
 It is assumed that only the donors are sensitive to the external potential and hereafter the term impurity  refers to the donor only.
 Exposing the impurities to the external harmonic potential compel the particles to accumulate around the potential minimum. Furthermore, applying a nonhomogeneous temperature profile in the vicinity of the potential minimum, may modify the external potential into a bistable effective potential as long as the condition given in Eq. (7)   is obeyed. Otherwise the effective potential is a monostable  potential.

Figure 1a gives  a plot of  the effective potential $V^{eff}$ versus $x$ for two different values of $\alpha$. The effective potential has two stable potential minima
at $\pm x_m$. Exploiting Eq. (6), one can see that the barrier height $\Delta V^{eff}$ and the width of the effective  potential $2x_{m}$ increase
as  $\alpha$ and  the trap depth $\Phi$  increase. On the contrary,  as the strength of the external potential $V_{0}$
and $\sigma$  decrease, the barrier height $\Delta V^{eff}$ and the width $2x_m$ increase.

We plot the dependence of the probability distribution $P_{ss}(x)$ as a
function of  $x$ in Fig. 1b. The figure depicts that the
impurities accumulate around the two potential wells. As $\Phi$
increases, the probability of finding the impurities around the
two stable points increases. When $\Phi$ increases, not only the
peak of   $P_{ss}(x)$ intensifies but also the distance between the two
peaks  $2x_{m}$ increases.
\begin{figure}[ht]
\centering
{
    \includegraphics[width=6cm]{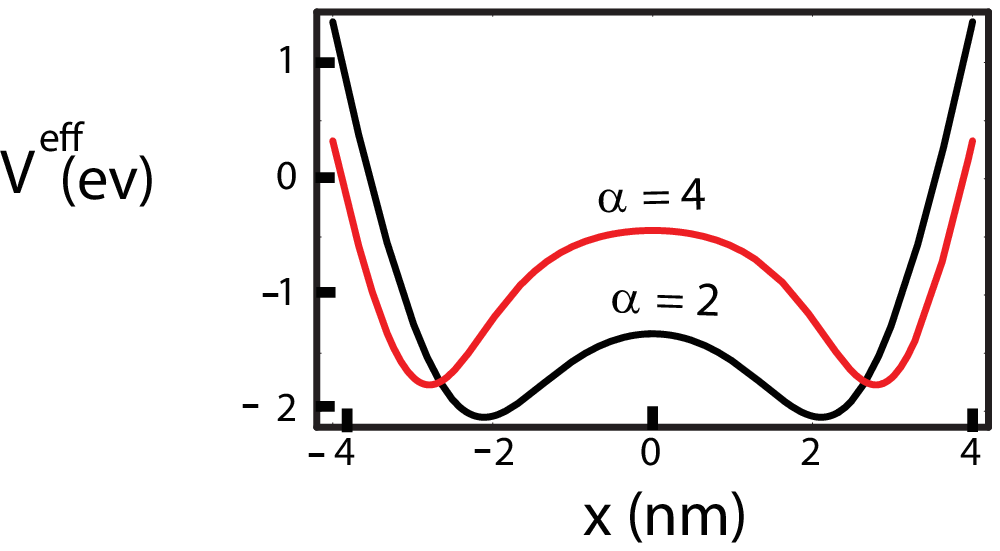}}
\hspace{1cm}
{
    \includegraphics[width=6cm]{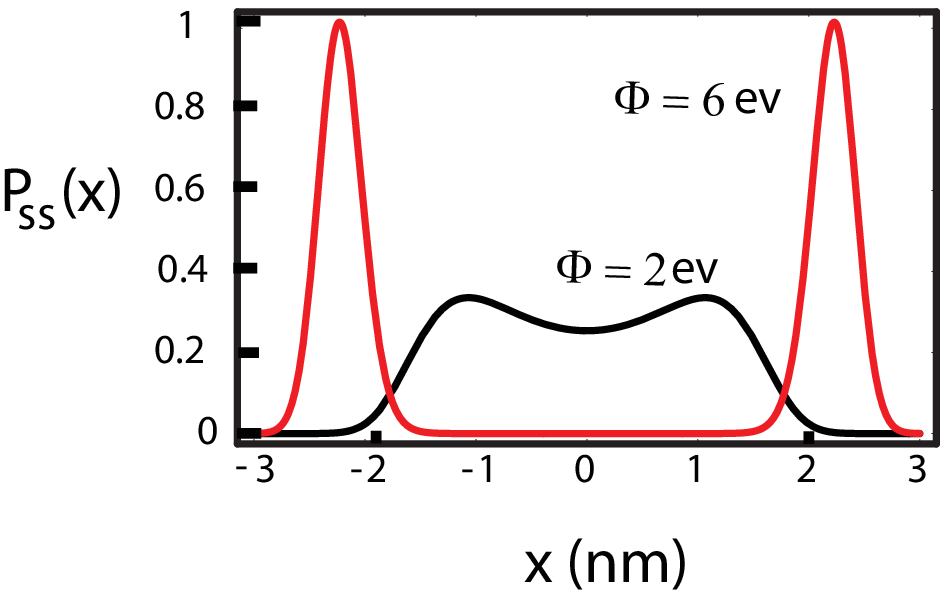}
}
\caption{(a) (Color online)  (a) The effective potential $V^{eff}(x)$ as a function of $x$ (nm)
 for parameters choice $\alpha=2$ (black line),  $\alpha=4$ (red line),
 $\Phi=4 eV$, $\sigma=1.1 nm$, $T_{c}=300${K} and  $V_{0}=0.4 {eV}/nm^{2}$. (b) Probability distribution $P_{ss}(x)$ as a function of $x$  for parameter choice
 $T_{c}=300${K},  $\alpha=1.5$, $V_{0}=0.4 {eV}/nm^2$ and  $\sigma=1.1 nm$. The black and red lines stand for $\Phi=2 {eV}$ and $\Phi=6 {eV}$, respectively. }
\label{fig:sub} 
\end{figure}

Let us now discuss how one can translocate the dopants to a desired
location along the semiconductor layer. By looking at how $V^{eff}$
and $P_{ss}(x)$ behave,  when the different model parameters vary, one
can infer the position of the impurities in the medium. As
discussed before when  $\alpha$ and $\Phi$ monotonously increase
or as $V_{0}$ and $\sigma$ constantly decrease, the  particles
which are accumulated  around the two potential minima leave the
central region $x=0$ and migrate to the peripheral part of the
semiconductor layer creating a depletion zone around $x=0$. This
suggests  ways of controlling  the
  {diffusivity} of the particles by tuning these parameters. Further increase in $\alpha$ or as one further  decrease
$V_{0}$ and $\sigma$,  the impurities  may become eliminated  from a certain region of the 
semiconductor layer. 

 {The impurities that are exposed  to the effective potential  exhibit noise-induced  nonequilibrium phase transition \cite{p1}   where the noise plays a counterintuitive role by inducing ordering phenomena.} Exploiting Eq. (7), one can see that   the regimes of  noise induced ordering phase transition   are demarcated by   the critical values of the different model parameters. The critical potential energy  $V_{0}^{*}$  and the critical trap depth $\Phi^{*}$ have a simple form:   $V_{0}^{*}=(\Phi\alpha/(2\sigma^2(\alpha+1))$ and   $\Phi^{*}=2V_{0}\sigma^{2}\left((\alpha+1)/ \alpha\right)$ while the other critical points are given by  $\sigma^{*}=\sqrt{(\alpha\Phi)/(2V_{0}(\alpha+1))}$ and $\alpha^*=(2V_{0}\sigma^2)/(\Phi-2V_{0}\sigma^2)$.
   When    $0<\Phi \leq \Phi^{*}$, $\alpha<\alpha^*$, $\sigma>\sigma^{*}$ or  $V_{0}>V_{0}^{*}$,  the effective potential is a monostable potential and   the  dopants  concentrate around the $x=0$ position. In such a case the impurity dynamics is similar to the previously  exposed works \cite{s1,s7}.   On the other hand when   $\Phi > \Phi^{*}$, $\alpha>\alpha^*$, $\sigma<\sigma^{*}$ or  $V_{0}<V_{0}^{*}$,  the effective potential has two stable minima and the impurities split into two impurity-rich {regions} which coexist with 
     impurity-poor {regions}. In this regime, the donors pile up towards the two potential minima while the oppositely charged particles hop in the opposite direction showing that a series of $p-n-p$ or $n-p-n$ junctions can be fabricated by manipulating the diffusion of the impurities. Note that the effect of internal field is neglected which is appropriate  at  low doping level.

Now let us study how  the thermally activated barrier crossing rate of the
impurities behaves.  Let us consider non-interacting impurities
initially situated at one of the potential minima. Due to thermal
fluctuation, the particles cross the potential barrier assisted by
the  thermal kicks they encounter along the reaction coordinate.  The crossing
rate for the impurities in  a high barrier limit $\Delta V^{eff}\gg
k_{B}T$ is approximated \cite{c26,c27} as
  \begin{equation}\label{eq:8}
        R = \sqrt{{|\omega_{0}| |\omega_{x_{m}}|\over 2\pi}} e^{\frac{-\Delta V^{eff}}{D}}.
 \end{equation}

 \begin{figure}[ht]
\centering
{
    \includegraphics[width=6cm]{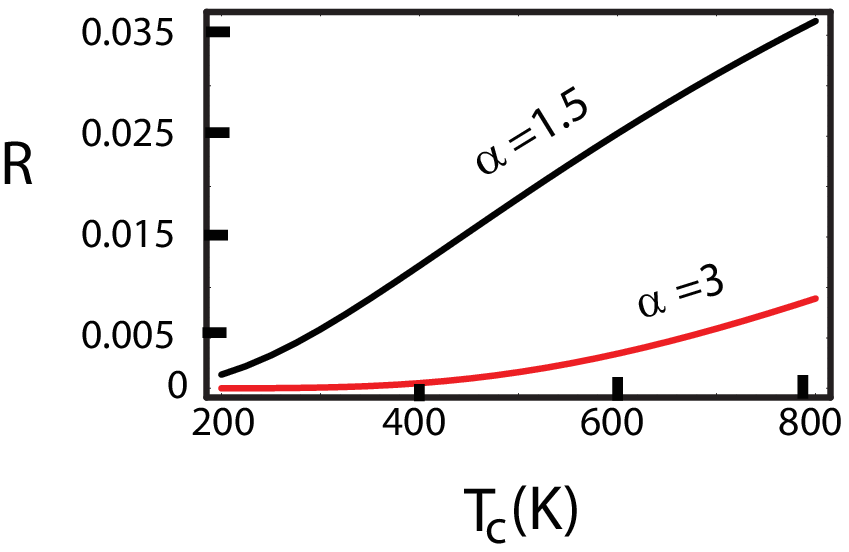}}
\hspace{1cm}
{
    \includegraphics[width=6cm]{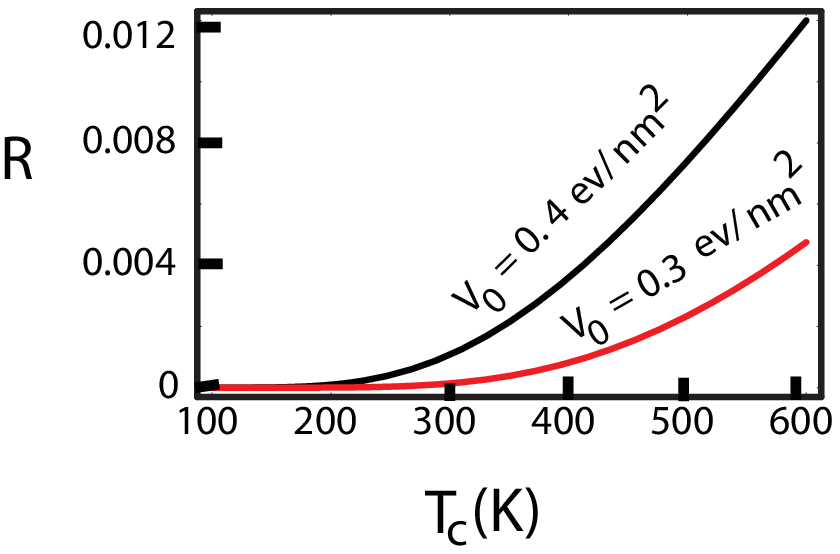}
}
\caption{(a) (Color online) Escape rate  $R$ versus $T_{c}$ in a unit of Kelvin for parameters choice $\alpha=1.5$ (black line),  $\alpha=3$  (red line),
 $\Phi=2 {eV}$ and  $\sigma=1 nm$. (b)The crossing rate $R$ versus $T_{c}$ in a unit of Kelvin for parameters choice  $V_{0}=0.4 {eV}/nm^2$ (black line)  and  $V_{0}=0.3 {eV}/nm^2$ (red line). Other parameters are fixed as   $\alpha=2$, $\Phi=2 {eV}$ and  $\sigma=1 nm$. }
\label{fig:sub} 
\end{figure} 
Figure 2a plots the {dependence} of the rate $R$ on $T_{c}$. The particles cross the  potential barrier at the expense of the thermal background kicks.
 Thus the rate increases with $T_{c}$. When $\alpha$ steps up,  $\Delta V^{eff}$ and $2x_{m}$ increase and  the particles attain difficulty in crossing
 the high barrier. Hence the rate becomes lower with $\alpha$.
Figure 2b  exhibits  that when $V_{0}$ rises, the escape rate
increases  due to the fact that increasing in $V_{0}$ results in a
lower barrier height.  Exploiting Eq. (13), one can also  see that the rate intensifies when $\Phi$ decreases or  as $\sigma$
increases.

\section{ Stochastic resonance of the impurities}

Let us now study  the noise-assisted dynamics for the  bistable
 potential  in the presence of time varying signal.  The interplay between noise and  time varying  force
in the bistable system may lead the system into stochastic resonance as long as
 the random kicks are adjusted in  an optimal way to the recurring external force.
Next  we study the dependence of the SR on the model parameters employing two state approximation.

Via two state model approach \cite{c12,c23}, two discrete states
$x(t) = \pm x_{m}$ are considered. Let us denote $n_+ $ and $n_-$
to be the probability  to find  the impurity in the right
($x_{m}$) and in the left ($-x_{m}$) sides of the potential wells,
respectively. In the presence of time varying  weak periodic force
(AC field) of type $v(t)= A_0 Cos (\Omega t)$, the master equation
that governs the time evolution of $n_{\pm}$ is given by
\begin{equation}
{\dot n_{\pm}(t)}=-W_{\pm}(t)n_{\pm}+W_{\mp}(t)n_{\mp}
\end{equation}
where $W_{+}(t)$ and $W_{-}(t)$ correspond to the time dependent
transition probabilities
 towards the right ($x_{m}$) and  the left ($-x_{m}$) sides of the
potential wells, respectively. The time dependent rate
\cite{c12,c23} takes a simple form
\begin{equation}
W_{\pm}=R \exp\left[\pm {x_{min} \over D}A_{0}\cos{(\Omega t)}\right],
\end{equation}
where $R$ is the Kramers rate for  the particles in the absence of
periodic force $A_0 = 0$.  For sufficiently small amplitude, one
finds the spectral amplification, $\eta$ to be \cite{c12,c28}
\begin{equation}
\eta=\left({x_{m}^2\over D}\right)^2{4R^{2}\over 4R^{2}+\Omega^{2}}.
\end{equation}

The input signal that modulates the symmetric
bistable system makes  one stable state  less stable than the other,
alternatively, over half  forcing period. When the random
switching frequency  matches the forcing angular frequency
 by tuning the noise intensity, the system
attains the maximum probability of escaping out of the less stable
state into the more stable one, before a random back switching
event takes place. Due to this reason, the particle is more likely
to be in the more stable state. Consequently  $\eta$ attains a
maximum value at a particular noise level $(D_{opt})$. Let us now
explore how the  $\eta$  behaves as a function of
the trap depth $\Phi$, {standard deviation} $\sigma$, applied potential energy  $V_{0}$ and
$\alpha$.

\begin{figure}[ht]
\centering
{
    \includegraphics[width=6cm]{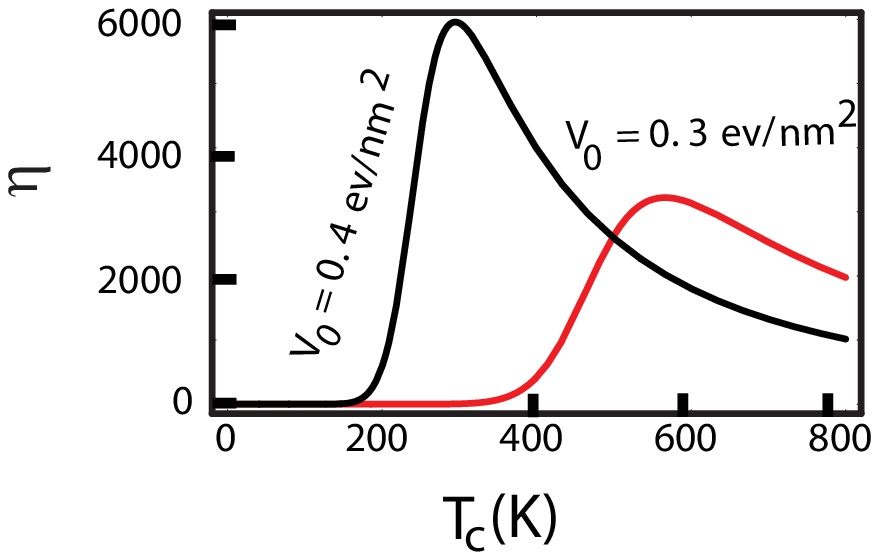}}
\hspace{1cm}
{
    \includegraphics[width=6cm]{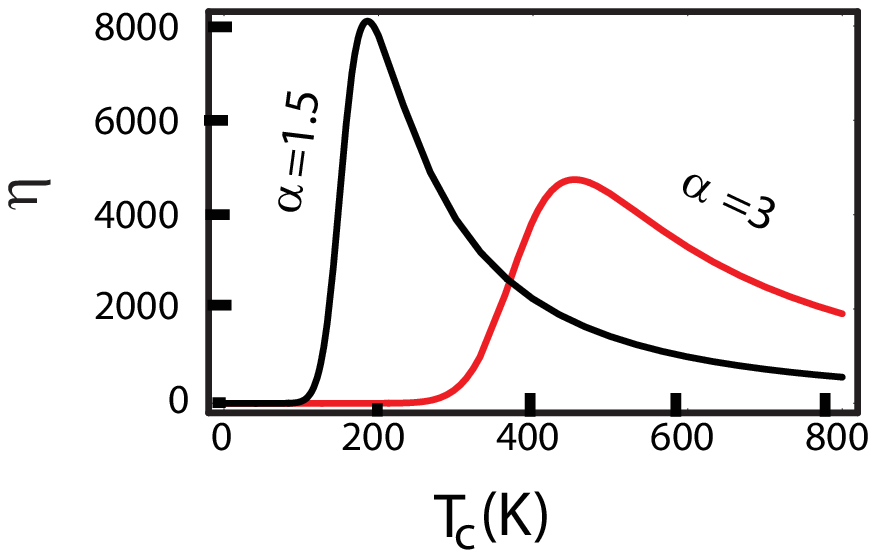}
}
\caption{(a) (Color online) Spectral amplification $\eta$ versus $T_{c}$ in a unit of Kelvin for parameters choice $V_{0}=0.4 {eV}/nm^2$ (black line) and $V_{0}=0.3 {eV}/nm^2$  (red line). Other parameters are fixed as   $\alpha=2$, $\Phi=2 {eV}$ and  $\sigma=1 nm$. (b) Spectral amplification $\eta$ versus $T_{c}$ in a unit of Kelvin for parameters choice $\alpha=1.5$ ( black line),  $\alpha=3$  (red line),
 $\Phi=2 {eV}$ and  $\sigma=1 nm$.}
\label{fig:sub} 
\end{figure}

Figure 3a plots the {dependence} of the $\eta$ on the noise strength
$T_{c}$ for parameter choice $V_{0}=0.4 eV/nm^2$ for black line
and $V_{0}=0.3 eV/nm^2$ for red line. Other parameters are fixed
as   $\alpha=2$, $\Phi=2 eV$ and  $\sigma=1 nm$. In the limit $T_{c}
\to 0$ and $T_{c}\to \infty$, $\eta \to 0$.  This is because when
the noise intensity $T_{c}$ is too small, the intrawell crossing rate
is too small while for large $T_{c}$ a similar loss of
synchronization occurs since the particles flip too many times
between the two stable points. In between $\eta$ attains an
optimal value at optimal $T_{c}^{opt}$. The peaks of $\eta$
increases with $V_{0}$. On the other hand,  $T_{c}^{opt}$ {increases}
as $V_{0}$ decreases. Figure 3b  exhibits the {dependence}  of
$\eta$ on $T_{c}$ for parameters choice $\alpha=1.5$ for black
line, $\alpha=3$ for red line, $\Phi=2 eV$,  $\sigma=1 nm$ and
$V_{0}=0.4 eV/nm^2$. When $\alpha$ {increases}, $T_{c}^{opt}$
increases; the peak of the  $\eta$ rises  when $\alpha$ diminishes. This  loss of  synchronization  with the
increase in $\alpha$ is  plausible because as   $\alpha$ {increases},
$\Delta V^{eff}$ and the width $2x_{m}$ increase;   the
particle crosses the high potential barrier at the expense of higher  thermal
strength $T_{c}$.

\begin{figure}[ht]
\centering
{
    \includegraphics[width=6cm]{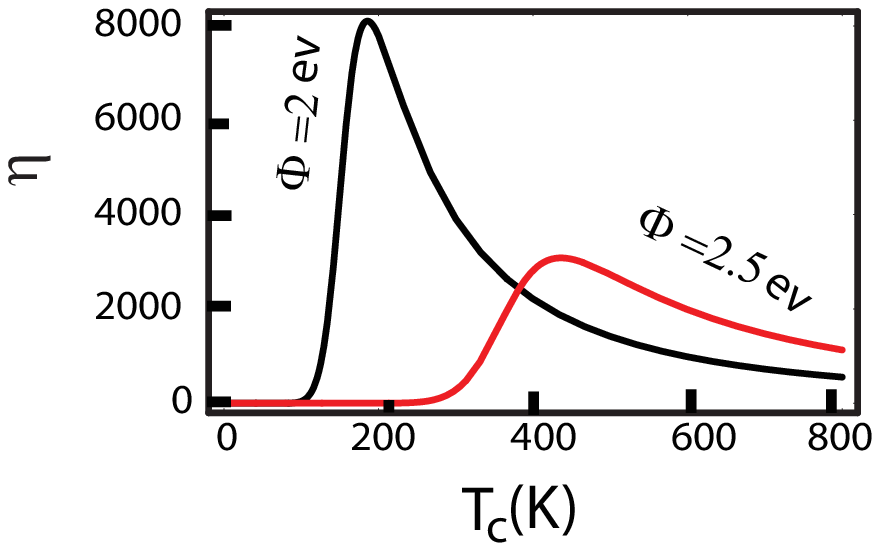}}
\hspace{1cm}
{
    \includegraphics[width=6cm]{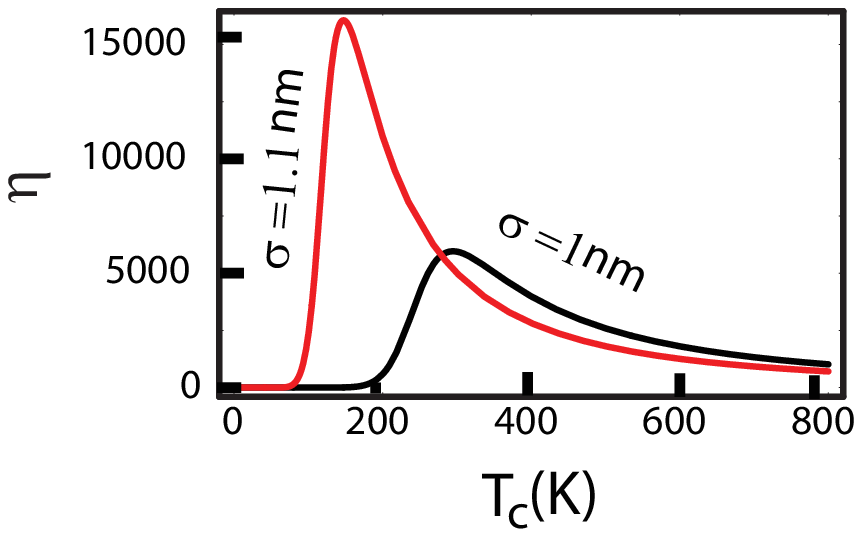}
}
\caption{(a) (Color online) Spectral amplification $\eta$ versus $T_{c}$ in a unit of Kelvin for parameters choice $\Phi=2 {eV}$ (black line) and $\Phi=2.5 {eV}$ (red line). Other parameters are fixed as   $\alpha=1.5$, $V_{0}=0.4 {eV}/nm^2$ and $\sigma=1 nm$. (b) Spectral amplification $\eta$ versus $T_{c}$ in a unit of Kelvin for parameters choice $\sigma=1$ (black line),  $\sigma=1.1$ (red line),
 $\Phi=2 {eV}$, $\alpha=2$ and  $V_{0}=0.4 {eV}/nm^{2}$. }
\label{fig:sub} 
\end{figure}

In Fig. 4a, we  show how the $\eta$  behaves as $\Phi$ varies.
When $\Phi$ increases, the peak of $\eta$ becomes smaller. $T_{c}^{opt}$
shifts to the right  as $\Phi$ {increases} showing  the response of signal to
the background noise strength is  significant at lower values of
$\Phi$. On the other hand, as $\sigma$ monotonously decreases,
the resonance becomes more  significant.  When $\sigma$  diminishes,
$T_{c}^{opt}$ increases (see Fig. 4b).

These results reveal the
weak signal passing through  the semiconductor layer can be
amplified and detected by tuning the temperature of the hot
locality  since the noise level can be directly affected by the
hot temperature. By tuning the angular frequency $\Omega$, one can
control the mobility of the impurities to a desired location. 
At the  resonance temperature 
$T_{c}^{opt}$, the impurities undergo a  fast unidirectional drift  from the  less stable 
potential minima to the more stable one    over half  forcing period 
revealing novel way of achieving a fast transportation of  impurities along the semiconductor {layer} 
without exposing the impurities to a higher temperature. 
Thus
the present study is crucial in the designing of artificial
semiconductor.

 \begin{figure}[ht]
\epsfig{file=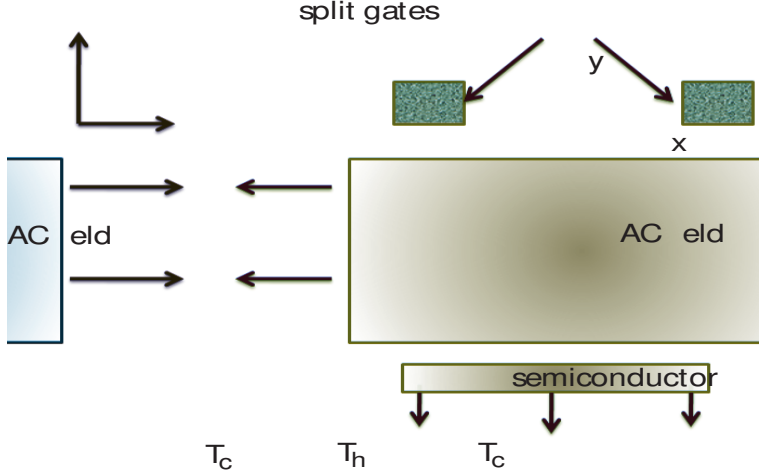,width=10cm} \caption{(Color online) Schematic diagram showing a 
semiconductor which is exposed to an  external potential (in $x$
direction) and   a nonhomogeneous temperature background, which is
hot around the potential minimum ($x=0$) and decays  to a {lower temperature}  as one goes away on both sides from the potential minimum.
The cross section along the $z$ direction is not shown. The impurities are assumed to be  uniformly distributed along the $y$ and  $z$ directions.
Furthermore, a time varying field (AC field) is applied along the
$x$ direction.
 }
\end{figure}

At this point we  stress that this theoretical work can be realized experimentally. As shown in
the schematic diagram (Fig. 5), one can model the harmonic
potential via the metallic gates which are kept at  a certain
voltage \cite{s1,s7}. The metallic gates  are situated  at the top
of the sample semiconductor. Similar to the work \cite{s1,s7}, the
harmonic potential is extended along the $x$ direction. A tiny
region around the potential minimum  $x=0$ is heated up and this
creates a depletion zone of the impurities  around $x=0$; the hot locality
forces the impurities to migrate towards the peripheral regions.
The diffusion of the particles can be controlled by tuning the
different model parameters. Applying periodic signals such  as  AC
field along the $x$ direction, modulates the effective potential.
With the proper adjustment of the different parameters, the system
may show the stochastic resonance and this arrangement could be used to amplify very weak signals.

\section{Summary and conclusion}

{This theoretical work exposes the way of manipulating the disfusiblity of the impurities along the semiconductor layer by locally heating the semiconductor layer around the potential minimum of the exerted   external harmonic potential. The theoretical results obtained in this work depicts that 
the dopant mobilizes to the peripheral regions when the trap depth $\Phi$
and $\alpha$ increase or when  $\sigma$ and the external
potential $V_{0}$ decrease.   The thermally activated
rate for the impurities  is  also studied at  high barrier limit
$\Delta V^{eff}\gg k_{B}T$. It is shown that the rate increases
with $V_{0}$  and   $\sigma$, and it  grows smaller when
  $\alpha$ and $\Phi$ increase.}

 {In the presence of periodic signals, the {dependence} of 
 the spectral amplification  $\eta$ on  the different control
 parameters  is explored. The peak of  $\eta$ rises  when  $\Phi$ and  $\alpha$ monotonously fall.  When  $\sigma$
 and $V_{0}$ increase,  the response of signal to the background temperature is significant.   The magnitude of optimum noise intensity
 $T_{c}^{opt}$ becomes considerable  with  $\Phi$ and   $\alpha$, and  when $V_{0}$  and   $\sigma$ grow smaller.}

{The model 
presented in
this work is not limited to   the impurities dynamics.
Rather  the present model is of
broader interest in various fields and    serves as a basic paradigm
in which to understand diffusion and   noise induced nonequilibrum
phase transition in  discrete systems. In conclusion, the proposed model is crucial in designing  artificial semiconductors. We 
believe that a semiconductor device under the suggested model can
be designed to detect weak signals of extremely small modulation.}

\section{Acknowledgment}

 MA  would like to thank Prof. W.
Sung for the interesting discussions he had during his  visit  at
APCTP, Korea. BA and MB would like to thank The International Programme in Physical Sciences, Uppsala University, Sweden for the support they have provided to our research group.

\end{document}